\documentstyle[prd,aps,psfig]{revtex}

\begin{document}
\draft
\twocolumn[\hsize\textwidth\columnwidth\hsize\csname
@twocolumnfalse\endcsname
\preprint{PACS: 98.80.Cq, IJS-TP-97/16,\\
{}~hep-ph/yymmnn}

\newcommand\lsim{\mathrel{\rlap{\lower4pt\hbox{\hskip1pt$\sim$}}
    \raise1pt\hbox{$<$}}}
\newcommand\gsim{\mathrel{\rlap{\lower4pt\hbox{\hskip1pt$\sim$}}
    \raise1pt\hbox{$>$}}}

\title{High Temperature Symmetry Breaking via Flat Directions}

\author{Borut Bajc$^{(1)}$ and Goran Senjanovi\'c$^{(2)}$}

\address{$^{(1)}${\it J. Stefan Institute, 1001 Ljubljana, Slovenia}}

\address{$^{(2)}${\it International Centre for Theoretical Physics,
34100 Trieste, Italy }}

\date{\today}
\maketitle

\begin{abstract}
We show that the natural presence of flat directions in 
supersymmetric theories allows for non-restoration of global 
and/or gauge symmetries. This has important cosmological
consequences for supersymmetric GUTs and in particular it 
offers a solution of the monopole problem.
\end{abstract}


\vskip1pc]

{\it A. Introduction}. \hspace{0.5cm} In spite of the everyday 
experience, it is well known that global symmetries 
may remain broken at temperatures much above the physical scale 
of the theory in question \cite{w74,ms79}. Although this has 
recently been confirmed by rigorous renormalization 
group improved methods \cite{roos96} and lattice 
calculations \cite{jl98}, it is not clear 
whether the same is true for the case of local symmetries 
\cite{bl95b,gprv98}. If one is willing to 
consider the universe with some large external charge then even 
local symmetries may naturally remain broken at high T 
\cite{lepton,charge}. This has important application for 
the fate of topological defects and in particular could solve the
cosmological problems of monopoles and domain walls
\cite{ds95,dms95,dms96}. For a recent review of these issues 
see \cite{s98}. 

However in what follows we are interested only in the generic high 
T behaviour of field theories without any external charge. It is 
known then at the level of a no-go theorem \cite{h82,m84} that in 
supersymmetric theories the phenomenon of non-restoration is not 
operative, even when nonrenormalizable interactions are included 
\cite{dt96,bms96,bs96}. This has unfortunate consequences not only 
for the monopole problem (which after all may be solved by inflation), 
but even more for the wrong vacuum problem of SUSY GUTs. Namely, 
these theories are normally characterized by the degenerate vacua 
at zero temperature, one of them being the unbroken symmetry one. 
If at high temperature the symmetry is restored, then the system 
would remain forever in this unphysical vacuum \cite{w82} for the 
barrier between the vacua is enormous (determined by the GUT scale). 

We wish to make our philosophy as clear as possible. Even if there 
is no phase transition, one needs to justify the initial 
condition of a homogeneous universe. For this reason we, as 
everybody else, assume that inflation took place at some point. 
This is indispensable for our program. However, one has no guarantee 
that inflation takes place after the production of topological 
defects or that the reheating temperature is below their masses. 
In fact, in grand unified theories this is often not the case. 
For this reason we believe that it is a must to look for other 
solutions to the problems of topological defects and the false vacuum. 
It is thus important to re-address the issue of high T behaviour of 
supersymmetric theories and this is precisely the aim of this letter. 

The crucial point in the no-go theorem of symmetry non-restoration 
in supersymmetry is based on the assumption of thermal equilibrium 
for all the fields of the theory. Whereas this is normally true, 
it is precisely the supersymmetric theories that may provide 
naturally a way-out of this assumption. Namely, these theories 
are generically characterized by a large number of flat directions, 
generically denoted by $\phi$ in what follows. Such flat directions, 
at least for large enough values of $\phi$ (see below), do not 
have strong enough interactions to be 
in thermal equilibrium and therefore the no-go theorem is not 
directly applicable to them \cite{dk98}. Remarkably enough, it 
can be shown that these flat directions may quite naturally 
possess vacuum expectation values much bigger than the temperature. 
This is the main point of our work, in full agreement with the 
recent results of Dvali and Krauss \cite{dk98}. In order to set 
the stage, we first briefly review the conventional situation of 
all the fields being in thermal equilibrium. 

\vspace{0.2cm}

{\it B. Supersymmetry in thermal equilibrium }. \hspace{0.5cm} 
The point here is quite simple. The fields being in thermal 
equilibrium have high temperature mass terms proportional to 
$T^2$. Now, in ordinary theories the mechanism of non-restoration 
is based on the possibility of negative dimensionless couplings 
in the scalar potential, thus allowing for the negative $T^2$ 
scalar masses. In supersymmetry, however, these couplings 
are the squares of the relevant Yukawa couplings and so can never 
be negative. This is roughly speaking the reason behind the 
no-go theorem for supersymmetry which can be rigorously proven 
under the assumption of thermal equilibrium and renormalizable 
couplings. The situation is somewhat more subtle in the 
nonrenormalizable case \cite{dt96}, but still the no-go theorem 
goes through \cite{bms96,bs96}. Essentially, the argument is as 
follows: if the main interaction is nonrenormalizable 
(say $|\phi|^6/M^2$, where $M$ is the large scale cutoff), its
sign in the potential must be positive in order for the potential 
to be bounded from below. This then immediately gives a positive 
temperature dependent mass term (in this case $T^4/M^2$). 

\vspace{0.2cm}

{\it C. Flat directions and symmetry non-restoration }. \hspace{0.5cm} 
As we already remarked, flat directions may not be in equilibrium 
at high temperature \cite{dk98}. This implies the absence 
of the $T^2$ mass term and paves the way for the possibility
of symmetry non-restoration. Now, at zero temperature 
one normally lifts the flat directions (at least the dangerous 
ones which break charge or color) by positive soft supersymmetry
breaking mass terms 
which push them to the origin. At high $T$, though, such soft terms 
are negligible compared to $T$ and since the temperature breaks
supersymmetry, it determines the stability point of flat 
directions. We will see that the run-away behaviour is quite
natural and can happen in a large class of supersymmetric 
field theories. The situation is reminiscent of the 
upside-down hierarchy of Witten \cite{w81}, when small supersymmetry
breaking terms may induce large vevs (even exponentially
large).
   
In order to illustrate this phenomenon, we describe it
on a simple toy model and then address the issue of realistic
gauge theories. Imagine a scalar superfield $q$ (it will
correspond to the matter superfields) with a renormalizable
self-interaction and a flat (or better almost flat) field
$\phi$, with the following superpotential

\begin{equation}
W = \lambda q^3/3 + W(\phi) \;,
\end{equation}   

\noindent
where we take

\begin{equation}
W(\phi)={\phi^{n+3}\over (n+3)M^n}\;\;\;(n\ge 1)\;,
\label{wphi}
\end{equation}

\noindent
so that in the limit $M\to\infty$ we have a completely flat 
direction. For this particular example the superpotential 
possesses a $U(1)$ R-symmetry

\begin{eqnarray}
\phi&\to&e^{i3\alpha}\phi\;,\nonumber\\
q&\to&e^{i(n+3)\alpha}q\;,\nonumber\\
W&\to&e^{i3(n+3)\alpha}W\;,
\end{eqnarray}

\noindent
and we wish to show that this symmetry is broken at high 
temperature. Of course, this symmetry has an anomaly, but 
this is of no importance to what we are trying to do. 

If $M$ is the Planck scale $M_{Pl}$, even the $n=1$ 
case may not suffice to bring $\phi$ in thermal equilibrium 
at high $T$, and in any case we imagine $n$ large enough to 
keep $\phi$ not in equilibrium (a reasonable assumption for 
what we call a flat direction). We shall quantify this more 
precisely below.

Now, for any reasonable $\lambda<1$ we will have a field $q$ 
in equilibrium and thus $q$ will run in thermal loops. Next, 
imagine that the K\" ahler potential is not trivial, 
exemplified by 

\begin{equation}
K(q,\phi)=q^\dagger q+\phi^\dagger\phi+
a{q^\dagger q\phi^\dagger\phi\over M^2}\;.
\end{equation}

Since we 
are interested only in the question of the vev of $\phi$ and 
since $\phi$ is not in equilibrium, we can take it to be a 
background field. Thus we can ignore its kinetic term and 
just concentrate on the field $q$. We first put its kinetic 
term in the canonical form by rescaling 

\begin{equation}
q\to (1+a|\phi|^2/M^2)^{-1/2}q\;.
\end{equation}

The potential at zero temperature is given by (before 
rescaling) 

\begin{equation}
V={\partial W\over\partial\phi^i}(K^{-1})^i_j
{\partial W^*\over\partial\phi_j^*}\;,
\end{equation}

\noindent
where $K^{-1}$ is the inverse of

\begin{equation}
K_i^j={\partial^2 K(\phi)\over
\partial\phi^i\partial\phi_j^*}\;.
\end{equation}

The relevant potential then simply becomes (for the 
rescaled field $q$)

\begin{equation}
V=\lambda^2{|q|^4\over (1+a|\phi|^2/M^2)^3}+
{|\phi|^{2(n+2)}\over M^{2n}}\;,
\end{equation}

\noindent
which can be expanded in $|\phi|^2/M^2$ to give 

\begin{equation}
V\approx \lambda^2|q|^4-3a\lambda^2|q|^4{|\phi|^2\over M^2}+
{|\phi|^{2(n+2)}\over M^{2n}}\;.
\label{int}
\end{equation}

Let us evaluate now the high temperature effective 
potential for the field $\phi$. Notice that there 
can be no term $T^2|\phi|^2$ due to the absence 
of $|q|^2|\phi|^2$ type terms in the potential 
(of course there is a $|q|^2T^2$ term which pushes 
the vev of $q$ to the origin). 
This is precisely the statement of $\phi$ being a 
flat direction and not being in thermal equilibrium 
($T^2$ mass terms can only arise for the fields in 
equilibrium). It does not mean though that the field 
$\phi$ does not feel the temperature at all. Clearly, 
from the $|q|^4|\phi|^2$ interaction in (\ref{int}), 
when $q$ is running in thermal loops, we will be left 
with a temperature dependent mass term for $\phi$. 

\begin{figure}[h]
\centerline{\psfig{figure=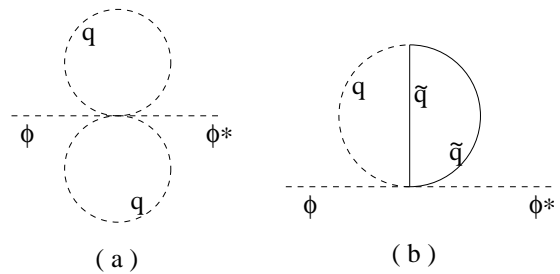,height=3.5cm,angle=-90}}
\centerline{   }
\caption{Diagrams which give the thermal mass to the 
flat direction $\phi$. Since $\phi$ is not in thermal 
equilibrium, only scalars $q$ and fermions $\tilde q$ 
are running in the loops.}
\end{figure}

Now, from the diagrams of Fig. 1 the high $T$ effective potential 
for $\phi$ can be easily evaluated \cite{bms96}

\begin{equation}
V_{eff}(\phi,T)=-{3a\lambda^2\over 32}{T^4|\phi|^2\over M^2}+
{|\phi|^{2(n+2)}\over M^{2n}}\;.
\label{veff}
\end{equation}

Clearly, for $a>0$ the $-$ sign in (\ref{veff}) guarantees a 
non-vanishing vev for $\phi$:

\begin{equation}
<\phi>^{n+1}=\left({3a\lambda^2\over 32(n+2)}\right)^{1/2}T^2M^{n-1}\;.
\label{vev}
\end{equation}

For $n=1$, $<\phi>\approx T$ and for large $n$, $<\phi>\gg T$ 
(we are by definition at $T\ll M$). Thus, not only does the 
field $\phi$ (an almost flat direction at $T=0$) have a vev at 
large $T$, but in general $<\phi>$ is expected to be much bigger 
than the temperature. For $a<0$, of course $<\phi>=0$. 
Thus the possibility of symmetry non-restoration depends 
crucially on the sign of the non-canonical piece in $K$. 

Let us now justify the natural assumption of $\phi$ being out 
of thermal equilibrium. From the interaction (\ref{wphi}) 
the mass of the $\phi$ particles is $m_\phi\approx <\phi>^{n+1}/M^n$, 
which, using the solution (\ref{vev}), becomes 

\begin{equation}
m_\phi\approx T^2/M\ll T\;.
\end{equation}

Thus, for $M=M_{Pl}$, which is what we expect in realistic 
cases, the effective decay rate for $\phi$ particles can be 
at most $m_\phi^2/T\approx T^3/M^2$ and is 
obviously much smaller than the expansion 
rate of the universe $H\approx T^2/M$. As expected, the field $\phi$ 
is not in thermal equilibrium. In exactly the same manner one 
shows that the K\" ahler interaction is also too weak to 
bring $\phi$ into the equilibrium.

We can generalize the above simple example by taking 
an arbitrary K\" ahler potential \cite{dk98}, in which 
case one has 

\begin{equation}
V(q,\phi)={\lambda^2|q|^4\over (K_q^q(\phi))^3}
+{\rm other\;\; terms}\;.
\end{equation}

As before, at high $T$, $|q|^4$ becomes proportional to 
$T^4$, so that clearly the vev for $\phi$ depends on the 
behaviour of $K_q^q(\phi)$: 
when it grows with $\phi$, this implies 
$<\phi>\ne 0$ (its precise value depends on whatever stabilizes 
the theory). Now, there is nothing unnatural about it and this 
is the central point of our work: not only is it possible to 
have symmetries broken at high $T$ in supersymmetric theories, 
but the presence of flat directions provides the most natural 
mechanism of symmetry non-restoration in general.

\vspace{0.2cm}

{\it D. Realistic case: gauge theories.} \hspace{0.5cm} The 
central ingredients, as we have seen, for the fields $\phi$ to 
develop a vev at high $T$ is supersymmetry breaking, provided 
by temperature, and a K\" ahler which grows with $\phi$. 
Supersymmetry must be broken in order for 
this to take place, and temperature plays 
this role naturally. What about the sign of $K_q^q(\phi)$ in 
realistic situations, so crucial for the phenomenon of non-restoration 
to take place? What about the heavy fields of the theory, those coupled 
to $\phi$? These important issues have been addressed and 
discussed at length in \cite{dk98}. For the sake of completeness 
we summarize here briefly their idea.

The answer to the second question is simple: any field which is 
coupled to $\phi$ and gets a mass from $<\phi>$, will necessarily 
decouple at temperatures $T\ll <\phi>$. As long as this is satisfied, 
such fields can be safely ignored, their contribution in thermal 
loops being suppressed by $exp(-<\phi>/T)$. The obvious example of 
such fields are the gauge bosons in the realistic case when $\phi$ has 
gauge interactions. 

The main concern must be addressed to the first question, i.e. the 
sign in $K_q^q(\phi)$. In general the light fields $q$ have 
both gauge interactions and Yukawa couplings to heavy states, and so 
in general the wave function renormalization will produce the 
logarithmic dependence in $\phi$ (through the heavy states masses) 

\begin{equation}
K_q^q(\phi)=1+(g^2(\phi)-h^2(\phi))log(\phi^2)
\end{equation}

\noindent
written symbolically ($h$ stands for the relevant Yukawa 
coupling). It is argued in \cite{dk98} that for sufficiently large 
gauge coupling, $\phi$ will necessarily have a large vev at 
high temperature. 

Now, what happens in a realistic theory such as the minimal 
supersymmetric standard model (MSSM)? As is well known, the flat 
directions are characterized by the holomorphic gauge invariant 
functions of the original superfields of the theory \cite{lt96}. 
In the MSSM with R-parity typical examples of such holomorphic 
invariants are 

\begin{equation}
lle^c\;\;,\;\; u^cd^cd^c\;\;,\ldots\;\;,
\end{equation}

\noindent
where $l$ is the leptonic doublet superfield and the rest are 
positron and anti-quark superfields. Of course at zero temperature 
these flat directions are lifted by soft supersymmetry breaking 
terms which drive the vevs of their scalar components to zero. 
These are the flat directions that we characterized above 
generically by $\phi$ and thus they are expected to have vevs 
at high temperature far away from the origin. If so, the 
electromagnetic gauge invariance will be broken at high 
temperature \cite{dk98} 
leading to the fast annihilation of monopoles \cite{lp80}, 
and thus solving the monopole problem \cite{dkl97}. 
Similarly there will be flat directions in the GUT extensions 
of the MSSM and the GUT symmetry may never be restored. In 
other words monopoles may not be created in the first place 
\cite{dk98,dkl97}. 

\vskip 0.2cm

{\it E. Summary and outlook.} \hspace{0.5cm} It should be clear 
from our work (we hope) that symmetry non-restoration at high 
temperature is a rather natural and generic phenomenon in 
supersymmetric theories. Simply, at high $T$ the flat directions 
may easily be lifted far from the origin due to the supersymmetry 
breaking induced by temperature. These flat directions may be the 
ones which for $T=0$ sit at the origin due to the soft supersymmetry 
breaking terms, or they may have large vevs even at $T=0$ as in the 
Witten's exponential hierarchy scenario \cite{w81}.

The large number of flat directions in the MSSM and supersymmetric 
GUTs would imply then a logical and natural possibility of the 
SM and GUT symmetries broken at high $T$. This offers the simplest 
and most natural solution to the monopole problem of GUTs. 

It could also provide a solution of another important cosmological 
problem of supersymmetric GUTs: the problem of the wrong vacuum 
(unbroken GUT symmetry vacuum). Namely, at $T=0$ these theories 
in general have discretely degenerate vacua with the SM gauge symmetry 
vacuum having the same ($=0$) energy as the unbroken GUT symmetry one. 
On the other hand, if at high $T$ symmetry were to be restored, this 
would make the theory get caught up in the unbroken vacuum 
forever, since the tunneling into the true SM vacuum is enormously 
suppressed due to the large barrier ($\approx M_X^4$) between the 
different vacua. If one could achieve the flat direction to be provided 
by the adjoint representation of the GUT symmetry group, then there may 
never be a troublesome phase transition and the broken vacuum would 
be preferred at all $T$. 

Notice that for the solution of the monopole problem the flat direction 
can be anything that provides the breaking of the GUT symmetry as to 
avoid the creation of monopoles. Equivalently, it could just be any 
flat direction which allows for the breaking of the electromagnetic 
gauge invariance in the MSSM at high $T$; this in turn allows for the 
annihilation of monopoles through the string-like flux tubes. 

\vskip 0.3cm

We would like to thank Gia Dvali, for stressing the importance 
of flat directions for high $T$ symmetry breaking, invaluable discussions 
and for sharing with us the preliminary results of his work with 
L. Krauss prior to publication. This work was supported 
by the Ministry of Science and Technology of Slovenia (B.B.) 
and by EEC under the TMR contract ERBFMRX-CT960090 
(G.S.). B.B. thanks ICTP for hospitality when part 
of this work was done.

\end{document}